\documentclass[fleqn,usenatbib]{mnras}

\usepackage[T1]{fontenc}

\DeclareRobustCommand{\VAN}[3]{#2}
\let\VANthebibliography\thebibliography
\def\thebibliography{\DeclareRobustCommand{\VAN}[3]{##3}\VANthebibliography}

\usepackage{graphicx}	% Including figure files
\usepackage{amsmath}	% Advanced maths commands
\usepackage{amssymb}	% Extra maths symbols

\title[Argon abundances in  Seyfert 2s]{Chemical abundances in  Seyfert galaxies -- VIII.   
 Argon abundance estimates}

\author[Monteiro \& Dors]{
A. F. Monteiro,$^{1,2}$\thanks{E-mail: adriano.santos@ifma.edu.br}
O. L. Dors,$^{1}$
\\
$^{1}$Universidade do Vale do Paraíba. Av. Shishima Hifumi, 2911, CEP: 12244-000, São José dos Campos, SP, Brazil\\
$^{2}$Instituto Federal do Maranhão. Av. Newton Bello s/n, CEP: 65906-335, Imperatriz, MA, Brazil
}

\date{Accepted XXX. Received YYY; in original form ZZZ}

\pubyear{2020}

\begin{document}
\label{firstpage}
\pagerange{\pageref{firstpage}--\pageref{lastpage}}
\maketitle

\begin{abstract}

For the first time,  the argon abundance relative to  hydrogen abundance (Ar/H) in the narrow line region of a sample of Seyfert~2 nuclei  has been derived.   In view of this,
optical narrow emission line intensities of a sample of 64 local Seyfert~2 nuclei ($z \: < \: 0.25$) taken from  Sloan Digital Sky Survey DR7 and  
measured by the MPA/JHU group were considered. 
   We adopted the $T_{\rm e}$-method for AGNs, which is based
 on direct determination of the electron temperature, together with a grid of
 photoionization model results, built with the {\sc Cloudy} code, to
 obtain a method   for the derivation of the Ar/H abundance. We find that
 for a metallicity range of $\rm 0.2 \: \la \: (Z/{\rm Z_{\odot}}) \: \la \: 2.0$,  Seyfert~2 nuclei present Ar/H abundance ranging from $\sim 0.1$ to $\sim 3$
 times the argon solar value,  adopting
 $\rm log(O/H)_{\odot}=-3.31$ and $\rm log(Ar/H)_{\odot}=-5.60$. These range of values correspond to 
 $\rm 8.0 \: \la \: (12+log(O/H) \: \la \: 9.0$ and 
 $\rm 5.4 \: \la \: (12+log(Ar/H) \: \la \: 6.9$, respectively.
 The  range  
 of Ar/H and Ar/O abundance values obtained  from our sample are in consonance with estimations from extrapolations
 of the radial abundance gradients to the central parts of the  disk for four spiral galaxies. We combined our abundance results with  estimates obtained from a sample of \ion{H}{ii} galaxies, which were taken from the literature, and found that the Ar/O abundance ratio decreases slightly as the O/H abundance increases.

\end{abstract}


%%%%%%%%%%%%%%%%%%%%%%%%%%%%%%%%%%%%%%%%%%%%%%%%%%

%%%%%%%%%%%%%%%%% BODY OF PAPER %%%%%%%%%%%%%%%%%%

\section{Introduction}
 Active Galactic Nuclei  (AGNs) present strong  metal emission-lines in their optical spectra, which when combined with hydrogen recombination
 lines, make it possible to estimate the abundance of heavy elements
and the metallicity in the gas phase of these objects.  AGNs play an  essential role in chemical abundance studies of nearby objects and of the early stages of  galaxy formation due to the aforementioned feature and to their high luminosity.

 Among the heavy elements, oxygen presents strong emission lines (i.e. [\ion{O}{ii}]$\lambda$3726, $\lambda$3729; 
 [\ion{O}{iii}]$\lambda$5007) emitted by its most abundant ions ($\rm O^{+}$,
 $\rm O^{2+}$) in the optical spectrum of gaseous nebulae (\ion{H}{ii} regions, Planetary Nebulae) and AGNs (e.g. \citealt{1978ApJ...223...56K, 1998AJ....116.2805V, Kennicutt2003, 2007A&A...463L..13M, 2015ApJS..217...12D, 2020MNRAS.496.2191F, Dors2020b}). Therefore,  
the total metallicity ($Z$) of the gas phase  from emission lines emitter objects is commonly traced by the oxygen abundance relative to hydrogen (O/H, e.g. \citealt{1991ApJ...380..140M, 2012MNRAS.422..215Y, Kewley2019}).
Other elements such as the noble gases  (e.g. Ne, Ar) present  emission-lines in the optical spectrum   emitted by only few of their ions (e.g. $\rm Ne^{2+}, Ar^{2+}$),  which make it necessary  to apply Ionization
Correction Factors (ICFs)  proposed by  \citet{1969BOTT....5....3P}  (see \citealt{2002astro.ph..7500S, 2014MNRAS.440..536D, 2013MNRAS.432.2512D, 2016MNRAS.456.4407D}) in order to account for the unobserved ions in the estimation of the total abundance.  The use of  ICFs  can introduce  uncertainties in order of 20\%  (e.g. \citealt{1996ApJ...458..215H, 1997AJ....114..713A, 2016ApJ...830....4C}) in the resulting total abundance
(for a detailed discussion on ICF uncertainties see \citealt{2014MNRAS.440..536D}).
Despite this drawback, noble gases can be used to derive the metallicity with some
advantages over oxygen, as they are also useful elements for determining constraints in  stellar nucleosynthesis studies.   Noble gas atoms can not combine in molecules formation  and they can not be trapped in dust
grains due to their quantum configuration, unlike the oxygen which is depleted onto dust in order of 0.1 dex
 (e.g. \citealt{Izotov2006, 2007MNRAS.376..353P}).

 The $T_{\rm e}$-method, which is based on direct estimation of the electron temperature, is widely used in the literature as the most reliable approach for determining the chemical abundance of heavy metals in gaseous nebulae (for a review see \citealt{Peimbert2017, Perez-Montero2017}).   The $T_{\rm e}$-method has been bolstered by the consonance between O/H abundance estimates in  \ion{H}{ii} regions in the solar vicinity and those obtained from observations of the weak interstellar \ion{O}{i}$\lambda$1356 line towards stars (see \citealt{pilyugin03} and references therein). Moreover, in the Milky Way and in nearby galaxies, good agreement between O/H abundance estimates in  \ion{H}{ii} regions and in B-type stars has recently been derived (e.g. \citealt{toribio17}). In this regard, abundance determinations of heavy metals (O, N, S, Ar, etc.) based on the $T_{\rm e}$-method have been carried out in   thousands   of  star-forming regions
 (SFs; i.e. \ion{H}{ii} regions, \ion{H}{ii} galaxies) at the local universe and for certain objects at high redshifts over decades (see \citealt{Dors2020b} and reference therein).

 Unfortunately, the situation is opposite for AGNs, where, except from oxygen, the majority of the abundances for other elements are not available in the  literature.
 In fact, the most complete  metal abundance determinations based on $T_{\rm e}$-method was carried out by \citet{1975ApJ...197..535O} for Cygnus~A ($z=0.05607$), who derived the O, N, Ne, S, and Fe abundances in relation to  hydrogen. After this pioneering work some few studies have applied the
$T_{\rm e}$-method  to abundance estimations in AGNs, however, in most cases, only 
oxygen abundance determinations   have been derived (e.g. \citealt{1992A&A...266..117A, 2008ApJ...687..133I,
2018ApJ...856...46R, 2018ApJ...867...88R, 2021ApJ...910..139R, Dors2015, Dors2020, Dors2020b}). Recently,  \citet{2020MNRAS.496.2191F}, adopting a methodology based on  a reverse-engineering of the $T_{\rm e}$-method, derived the first (N/O)-(O/H) relation for AGNs.  Although
studies relied on
photoionization models have been applied to derive metal abundance in AGNs (e.g. \citealt{grazina84, ferland86,  Storchi-Bergmann1998, groves2006emission, Feltre2016, Castro2017,  Perez-Montero2019, Thomas2019, Carvalho2020, 2021MNRAS.501.1370D, 2021MNRAS.505.4289P}),   most of them  have produced only estimations for O/H or metallicity. 
Since the electron temperatures throughout the emission nebula are computed by thermal balancing (see the seminal paper by \citealt{1967ApJ...147..556W}), the abundances of the most important elements are used as input parameters in photoionization models. All the lines, even if not observed, contribute to the gas cooling rate. In most of the papers which describe the results obtained  by using photoionization models, the ele\-ment relative abundances to H,  which were not found to be particularly different  from the solar ones, are  unfortunately not published in the literature.

 The use of photoionization model to derive 
abundance of different elements other than the oxygen (e.g. N, S, Ar) is (relatively) difficult, hence, it is necessary to find a solution for the electron temperature (or for O/H, the main cooler element) and for the ionization degree of the gas. Afterwards, the lines of the element under study must be adjusted in order to obtain its abundance (see, for instance, \citealt{2007MNRAS.377.1195P, 2010MNRAS.404.2037P, 2017MNRAS.469.3125C, 2017MNRAS.471..562C, Dors2017, 2019A&A...622A.119P, 2021MNRAS.501.1370D}). This procedure can produce a degeneracy among  nebular parameters, resulting in somewhat uncertain  elemental abundances \citep{2016A&A...594A..37M, 2018cagn.conf...33M}. In this sense, the use of the $T_{\rm e}$-method
produces more exact elemental abundance values  in comparison with those estimated through photoionization models.

In particular, the argon abundance determination in AGNs is very important in the study of galaxy evolution and stellar nucleosynthesis, since the stellar production and later ejection of this element to the Interstellar Medium (ISM) 
 in the high metallicity regime can be accessed.
The stellar nucleosynthesis theory predicts a primary origin for oxygen and argon (also for sulphur and neon) which are predominantly produced
on relatively short timescales by core-collapse supernovae
(SNe; massive stars) explosions (e.g. \citealt{1995ApJS..101..181W}). 
Thus, assuming a universal Initial Mass Function (IMF)\footnote{For a discussion on the universality of the IMF see, for example, \citet{2010ARA&A..48..339B}.},  there is expectation for a relatively constant value of the Ar/O abundance ratio with the O/H (or metallicity) variation, as derived by several authors in chemical
abundance studies of   SFs (e.g. \citealt{1995ApJ...445..108T, 1999ApJ...511..639I, 2006ApJ...636..214V, 2011A&A...529A.149G, 2006ApJ...636..214V}). However, some authors have found different behaviour
of Ar/O with O/H. For example, \cite{Izotov2006}, who used a large sample of
star-forming galaxies whose observational data were taken from 
SDSS-DR3 \citep{2005AJ....129.1755A}, found that Ar/O abundance ratio decreases by 0.15 dex  with the increase of O/H for the range $\rm 7.1 \: \la \: 12+\log(O/H) \: \la \: 8.5$ (see also \citealt{Perez-Montero2007}).
On the other hand,  recent results from the CHAOS project \citep{2015ApJ...806...16B} derived by \citet{2020ApJ...893...96B}, who applied the $T_{\rm e}$-method to 190 individual \ion{H}{ii} regions located in nearby galaxies, 
found Ar/O about constant and similar to the solar value for the range $\rm 8.3 \: \la \: 12+\log(O/H) \: \la \: 9.0$, and a high decrease of this abundance ratio for the very low abundance regime. Finally, \citet{Kennicutt2003} hinted that the Ar/O abundance decreases at high metallicity
($\rm 8.5 \: \la \: 12+\log(O/H) \: \la \: 8.7$) in  the M\,101 spiral galaxy.

 For the very high metallicity regime   ($\rm 12 +\log(O/H) \: \ga \:  8.8$), the behaviour of Ar/O with O/H is poorly known as well as its abundance in AGNs. Recently, \citet{Dors2020b} adapted the 
$T_{\rm e}$-method to chemical abundance studies of AGNs,  which made it  possible to obtain direct O/H estimates up-to  the very high metallicity regime, 
i.e. $\rm 12+log(O/H)\approx9.3$, an abundance value which is about 0.3 dex higher
than the maximum value obtained for SFs \citep{2007MNRAS.376..353P, 2020ApJ...893...96B}. Therefore, abundance studies in AGNs through the  $T_{\rm e}$-method
allow for the calculations of reliable Ar/H abundances in this class of object
and  further investigating  the Ar/O-O/H relation at very high metallicity regime, which is inaccessible in SF abundance studies.

In this context, we used the $T_{\rm e}$-method and 
 developed a new methodology based on photoionization model
 to calculate the abundance of  the argon relative to hydrogen in the narrow-line regions  (NLRs) of Seyfert 2 galaxies, whose data were taken from
 the SDSS-DR7 \citep{2000AJ....120.1579Y}. The present study is organized as follows.   In Section~\ref{method}, 
the  observational data and the methodology
used to estimate the oxygen and argon abundances are presented.
The results   and 
the discussion are presented in Sect.~\ref{resc}. Finally, the conclusion of the outcome is given in Sect.~\ref{conc}.

\section{Methodology}
\label{method}

To determine the total abundance of the argon and oxygen in relation to  hydrogen abundance (Ar/H, O/H),
firstly, we consider optical emission-line intensities of AGNs type Seyfert~2
from Sloan Digital Sky Survey Data Release 7 (SDSS-DR7, \citealt{2000AJ....120.1579Y}).
These observational data were used to calculate the  Ar/H and O/H abundances using the $T_{\rm e}$-method. We used photoionization models, built with the {\sc Cloudy} code \citep{Ferland2017}, in order to obtain an ICF for the $\rm Ar^{2+}$   and an estimation of the temperature  for the gas region occupied by this ion. In what follows, a description of the observational data
and the methodology adopted to obtain the abundances are presented.

\subsection{Observational data}
\label{obs}
We used optical ($3000 \: < \:\lambda($\AA$) \: < \: 7200$) 
reddening-corrected emission-line intensities of a sample  of Seyfert 2 nuclei obtained from the SDSS-DR7 \citep{2000AJ....120.1579Y} 
data made available by MPA/JHU group\footnote{https://wwwmpa.mpa-garching.mpg.de/SDSS/DR7/}. The sample consists of 463 Seyfert~2 nuclei with redshift 
$z \: \la \: 0.4$ and stellar masses of the hosting galaxies in the range of 
$9.4 \: \la \:  \log(M/{\rm M_\odot}) \:  \la \: 11.6$ selected by
 \citet{Dors2020}.  From this sample, we  
 considered a sub-sample containing only objects which have the
 [\ion{O}{ii}]$\lambda$3726+3729, [\ion{O}{iii}]$\lambda$4363, [\ion{O}{iii}]$\lambda$5007, H$\alpha$, [\ion{S}{ii}]$\lambda$6716, 
[\ion{S}{ii}]$\lambda$6731  and [\ion{Ar}{iii}]$\lambda$7135 emission-lines
measured with an error lower  than 50\,\%.  This criterion reduced the sample to  64 objects out of the 463 selected by \citet{Dors2020}, with redshif in the range of
$0.04 \: \la \: z \: \la \: 0.25$ and range of the stellar masses of the hosting galaxies
$9.9 \: \la \:  \log(M/{\rm M_\odot}) \:  \la \: 11.2$.

\begin{figure}
	\centering
	\includegraphics[width=\linewidth]{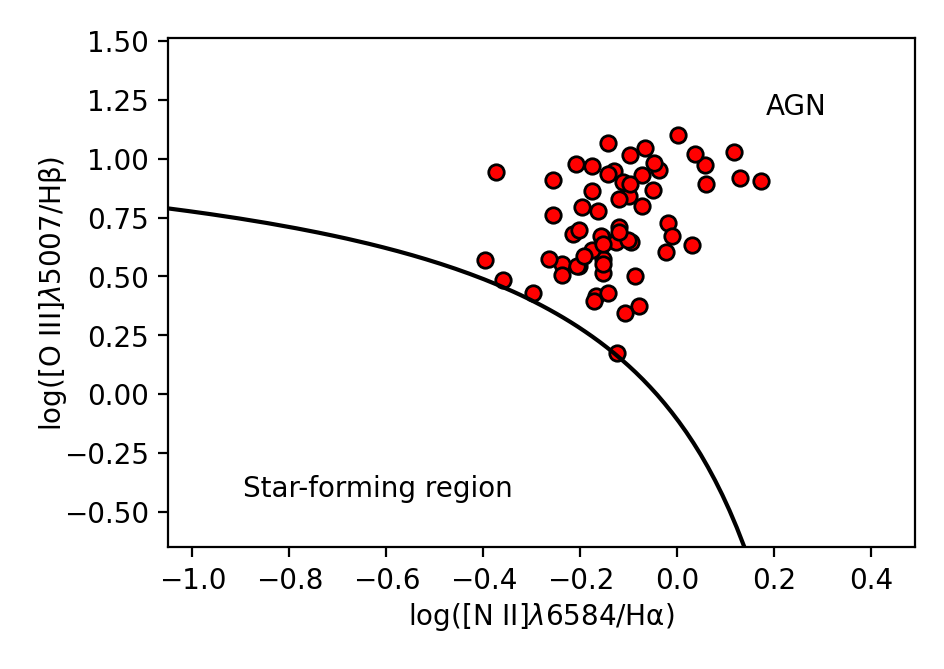}
	\caption{Diagnostic diagram log([\ion{O}{iii}]$\lambda$5007/H$\beta$) versus log([\ion{N}{ii}]$\lambda$6584/H$\alpha$). Red points 
	represent our sample of Seyfert~2 nuclei (see Sect.~\ref{obs}) whose
	observational emission-line ratios were taken from the SDSS-DR7 \citep{2000AJ....120.1579Y}
	and measured by the MPA/JHU group. The solid black line represents the AGN/Star-forming region  separation line proposed  by \citet{Kewley2001} and given by the  Equation~\ref{kewline}}.
	\label{fig:bptsample}
\end{figure}

In Figure~\ref{fig:bptsample}, a standard 
Baldwin, Phillips \& Terlevich diagram \citep{1981PASP...93....5B}
 log([\ion{O}{iii}]$\lambda$5007/H$\beta$) versus log([\ion{N}{ii}]$\lambda$6584/H$\alpha$), the observational line ratio intensities for the 64 objects are represented by red points. Also in this figure, the theoretical classification  criterion 
(represented by the black line)
proposed by \citet{Kewley2001}, which depict that objects with 
\begin{equation} 
\label{kewline}
\rm log([O\:III]\lambda5007/H\beta) \: > \: \frac{0.61}{log([N\:II]\lambda6584/H\alpha)-0.47}+1.19
\end{equation}
are classified as AGNs, otherwise, as  SFs,
is shown. It can be seen
that the sample cover a large range of ionization degree
and metallicity
 hence a wide range of [\ion{O}{iii}]/H$\beta$ 
and [\ion{N}{ii}]/H$\alpha$
are observed (e.g. \citealt{groves2006emission, Feltre2016, Carvalho2020}).

\begin{figure}
    \centering
    \includegraphics[width=\linewidth]{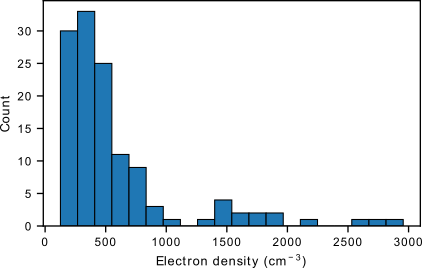}
    \caption{Histogram showing the distribution of
    electron density values from our sample of 64 objects (see Sect.~\ref{obs})   calculated with PyNeb routine \citep{Luridiana2015} and considering an electron  temperature of $10^4$ K. The y-axis represents the
    number of object with a given 
    range of electron density value.}
    \label{fig:hist_dens}
\end{figure}

The electron density ($N_{\rm e}$) of each one of the 64 Seyfert~2 nuclei was calculated from  the $[\ion{S}{ii}]\lambda 6716/\lambda 6731$ line ratio, assuming an electron temperature of 10\,000 K, and using the {\sc PyNeb} routine \citep{Luridiana2015}.
  In Fig.~\ref{fig:hist_dens} a histogram with
the $N_{\rm e}$ distribution of our sample is shown.
It can be seen that the $N_{\rm e}$ values for our sample (the maximum value is
about 3000 cm$^{-3}$) are lower than the
critical density (i.e. $10^{4-8} \rm \: cm^{-3}$, see \citealt{Vaona2012}) 
of the emission lines involved in the present study, therefore,
effects of collisional de-excitation are negligible in our abundance estimates.
In \cite{Dors2020},  a complete description of the selection criteria  
adopted to obtain the sample as well as a discussion about aperture effects on the abundance determination is presented. Moreover, effects of electron density variation along the AGN radius, X-Ray dominated regions,  shock and electron temperature fluctuations
 in abundance determinations have been discussed by \citet{Dors2020b} and
\citet{2021MNRAS.501.1370D} and these  are not repeated here.

\subsection{$T_{\rm e}$-method}
\label{secteme}

  It was possible to estimate
the Ar/H and O/H abundances through the $T_{\rm e}$-method for the sample of 64 objects.  In view of the $T_{\rm e}$-method, we followed a similar  methodology proposed by \citet{Perez-Montero2017} and  \citet{Dors2020b}.

\subsubsection{Oxygen abundance}

First, for each object, we calculated the temperature of the high ionisation   gas zone ($t_{3}$) and the electron density ($N_{\rm e}$) based on the dependence of  these nebular parameters  on the
[\ion{O}{iii}]($\lambda4949$+$\lambda5007$)/$\lambda4363$ 
and [\ion{S}{ii}]$\lambda6716$/$\lambda6731$ line ratios, respectively.  We used the
 function {\sc getCrossTemDen} from  the {\sc PyNeb} code \citep{Luridiana2015},
 where the value of each parameter  was obtained by
interacting over the two sensitive line ratios above. The errors in
$N_{\rm e}$ and $t_{3}$ were calculated  adding a Monte Carlo random-gauss values to the sample with the function {\sc addMonteCarloObs} in the {\sc PyNeb} code.
The symbol $t_{3}$ represents the electron temperature in units
of $10^4$ K.
The mean value derived for  $t_{3}$ from our sample is 
$\sim1.5$ and the uncertainty is $\sim0.3$. For the electron density, we found a mean  value of $470 \: \rm  cm^{-3}$ and an uncertainty of $\sim 250 \: \rm cm^{-3}$.

Since it is not possible to estimate the temperature for the  low 
 ionization  gas zone ($t_{2}$) due to the absence of the  auroral [\ion{N}{ii}]$\lambda$5755  and
[\ion{O}{ii}]$\lambda$7319, $\lambda$7330 line intensities,
the following theoretical relation  (Eq.~\ref{t2t3new}) between $t_{2}$-$t_{3}$, derived
by \citet{Dors2020b} for chemical abundance studies of AGNs and obtained
by using a photoionzation model grid built with the {\sc Cloudy} code 
\citep{Ferland2013} by \citet{Carvalho2020},  was adopted:
\begin{equation}
\label{t2t3new}
t_{2}=({\rm a} \times t_{3}^{3})+({\rm b} \times t_{3}^{2})+({\rm c} \times t_{3})+{\rm d},
\end{equation}
where $\rm a=0.17$, $\rm b=-1.07$, $\rm c=2.07$ and  $\rm d=-0.33$,
while $t_{2}$ and $t_{3}$ are in units of $10^{4}$ K.
In \citet{2021MNRAS.501L..54R} this  theoretical $t_{2}$-$t_{3}$ relation   was
compared with direct electron temperature estimations, calculated  through  observational
auroral emission lines for a sample of  AGNs,   and a good agreement was found between
them.  However, these authors indicated  that some cautions must be taken into account in the use of
Eq.~\ref{t2t3new} for AGNs with strong   outflowing gas.

 To calculate the ionic abundances of O$^{2+}$ and O$^{+}$ relative to H$^{+}$  the expressions provided by \citet{Perez-Montero2017}:
\begin{multline}
12+\log\left( \frac{\mathrm{O}^{2+}}{\mathrm{H}^{+}}\right) = \log \left( \frac{1.33 \times I([\ion{O}{iii}]\lambda5007)}{I(\mathrm{H\beta})} \right) + 6.1868 \\ + \frac{1.2491}{t_3} - 0.55816\log(t_3)
\end{multline}
and
\begin{multline}
12+\log\left( \frac{\mathrm{O}^{+}}{\mathrm{H}^{+}}\right) = \log \left( \frac{I([\ion{O}{ii}]\lambda3726+\lambda3729)}{I({\rm H}\beta)} \right) + 5.887 \\ + \frac{1.641}{t_2} - 0.543\log(t_2) + 0.000114\, n_{\rm e},
\end{multline}
respectively, where $n_{\rm e}$ is the electron density  $N_{\rm e}$ in units of 10$^{4}$ cm$^{-3}$, were used.
The uncertainty of $t_{3}$ was considered in the determination of $t_{2}$. Thus,
in the derivation of  $\rm O^{2+}/H^{+}$ ionic abundance the uncertainties
of the [\ion{O}{iii}]/H$\beta$ line ratio and  of $t_{3}$ were taken into account. The same
procedure was considered for the derivation of  $\rm O^{+}/H^{+}$, where the
$n_{\rm e}$ and $t_{2}$ uncertainties were also taken into account.

Finally, the total abundance of the oxygen is approximated by
\begin{equation}
\label{eqt6}
{\rm
\frac{O}{H}=ICF(O)\: \times \: \left[\frac{O^{2+}}{H^{+}}+\frac{O^{+}}{H^{+}}\right]}. 
\end{equation}
It is not possible to derive ICF(O),  for instance through
the expressions proposed by \citet{1977RMxAA...2..181T}
and \citet{Izotov2006}, due to  the
absence of the \ion{He}{ii}$\lambda$4686  emission line in our observational data.
 Therefore, we assumed ICF(O)=1.20, an average value obtained by 
\citet{2020MNRAS.496.2191F} and \citet{Dors2020b}.
 Oxygen ICF values translates into an abundance correction for AGNs
of $\sim$0.1 dex,  which is in agreement with the uncertainty derived in \ion{H}{ii} regions via $T_{\rm e}$-method estimates (e.g. \citealt{Kennicutt2003, Hagele2008, 2020ApJ...893...96B}).

\subsubsection{Argon abundance}

Generally, to estimate the temperature of the electrons   excitating the
$\rm Ar^{2+}$, the approach
proposed by \citet{Garnett1992} which is based on photoionization model results is assumed in chemical abundance studies of  SFs
(e.g. \citealt{Kennicutt2003, Hagele2008}).   A similar electron temperature is assumed for the gas regions
where the $\rm Ar^{2+}$ and $\rm S^{2+}$ ions are located, i.e.
\begin{equation}
\label{eqgarne}
T_{\rm e}(\ion{Ar}{iii})= T_{\rm e}(\ion{S}{iii}).    
\end{equation}
 The $T_{\rm e}(\ion{S}{iii})$ can be calculated directly from   [\ion{S}{iii}]($\lambda9069$+$\lambda9532$)/$\lambda6312$
or through the following theoretical
relation proposed by \citet{Garnett1992}, i.e. 
\begin{equation}
\label{teqs3o3}
T_{\rm e}(\ion{S}{iii})=0.83 \:  T_{\rm e}(\ion{O}{iii}) + 1700 \rm \: K
\end{equation}
when  [\ion{S}{iii}]($\lambda9069$+$\lambda9532$)/$\lambda6312$ line ratio can not be measured. In Eq.~\ref{teqs3o3}, 
$T_{\rm e}(\ion{O}{iii})= t_{3} \: \times \:10^{4}\:\rm K$.

\begin{figure}
\centering
\includegraphics[width=0.9\linewidth]{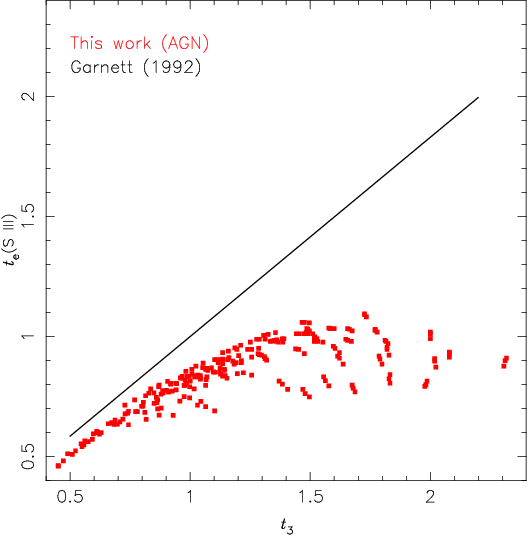}
\includegraphics[width=0.9\linewidth]{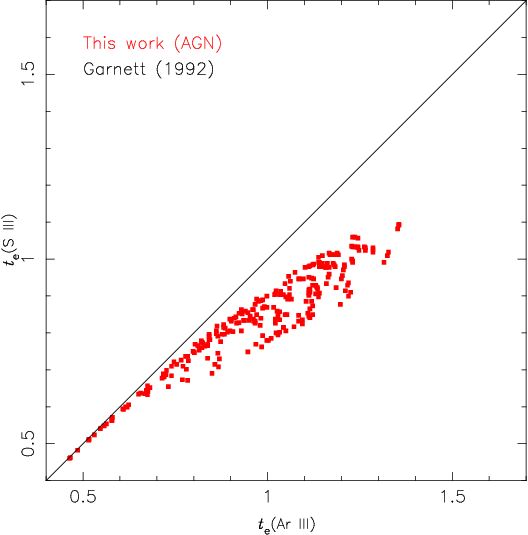}
\caption{Temperature values for the $\rm S^{2+}$, $\rm Ar^{2+}$
and $\rm O^{2+}$
predicted by the photoionization models built
by \citet{Carvalho2020} by using the {\sc Cloudy} code \citep{Ferland2013}. 
The values correspond to the model predicted  mean temperature (in units of $10^4$ K) for each ion   over the nebular AGN radius  times the electron density.
 Bottom panel: temperature values for the $\rm S^{2+}$ versus 
those for $\rm A^{2+}$. The black line corresponds to
equality between the temperatures (Eq.~\ref{eqgarne})
 proposed by \citet{Garnett1992} for \ion{H}{ii} regions. Red points represent photoionization model results. Top panel: same as the bottom panel but for $\rm S^{2+}$ versus $t_{3}$, where $t_{3}$ represents the
 temperature for $\rm O^{2+}$. Black line represents Eq.~\ref{teqs3o3}
 proposed by \citet{Garnett1992}.}
\label{far3s3}
\end{figure}

However, \citet{Dors2020b} showed that, in general, AGNs
present a distinct electron temperature structure than
\ion{H}{ii} regions (see also \citealt{2021MNRAS.501.1370D, 2021MNRAS.501L..54R, 2021arXiv210904596A}).
Thus, it is worthwhile to ascertain the validity of Eqs.~\ref{eqgarne} and \ref{teqs3o3} for AGNs. In view of this, we used the results
  from the photoionization model grid by \citet{Carvalho2020} and 
considered predictions for $T_{\rm e}(\ion{A}{iii})$ and  $T_{\rm e}(\ion{S}{iii})$. This grid of photoionization models  considered
a wide range of AGN nebular parameters whose optical predicted emission lines reproduce those of a large sample of Seyfert~2 nuclei (for a detailed 
description of these models see \citealt{Carvalho2020}). The temperature values
predicted by the photoionization models and considered here correspond to   the mean temperature for  $\rm Ar^{2+}$ and $\rm S^{2+}$ over the nebular AGN radius  times the electron density.
In Fig.~\ref{far3s3}, bottom panel, the photoionization
model predictions for $T_{\rm e}(\ion{S}{iii})$ versus $T_{\rm e}(\ion{Ar}{iii})$
(in units of $10^4$ K) and the approach given by Eq.~\ref{eqgarne} 
are shown. It can be seen that, in contrast to \ion{H}{ii} regions, $T_{\rm e}(\ion{A}{iii})$  is generally higher   than $T_{\rm e}(\ion{S}{iii})$, indicating that Eq.~\ref{eqgarne}
is not valid for AGN abundance studies.
The same can be seen
in Fig.~\ref{far3s3}, top panel, where the AGN model predictions
show a large deviation from the temperature relation given by 
Eq.~\ref{teqs3o3}. 

\begin{figure}
\centering
\includegraphics[width=0.9\linewidth]{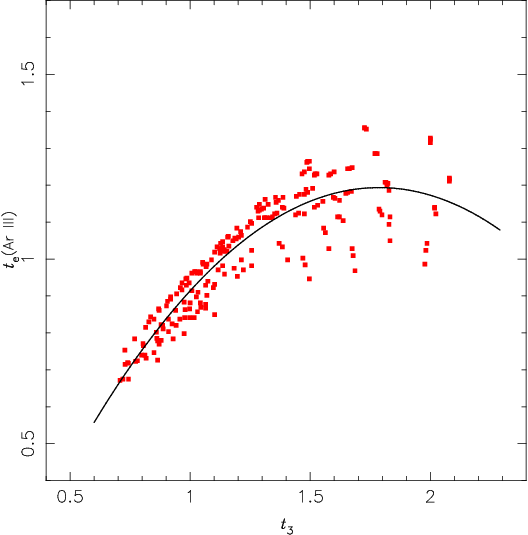}
\caption{Same as Fig.~\ref{far3s3} but for $T_{\rm e}(\ion{Ar}{iii})$
versus $t_{3}$. Black curve represents a fit to the points
represented by Eq.~\ref{fitaro}.}
\label{far3o}
\end{figure}

In order to estimate a more precise temperature to derive the $\rm Ar^{2+}/H^+$   ionic abundance, the photoionization models by \citet{Carvalho2020} were used to obtain a relation between 
$T_{\rm e}(\ion{Ar}{iii})$ and $t_{3}$ as shown in Fig.~\ref{far3o}  because $t_{3}$ can be derived
directly from [\ion{O}{iii}]($\lambda4959+\lambda5007$)/$\lambda4363$
observational line ratio of our sample. In Fig.~\ref{far3o},
$t_{\rm e}(\ion{Ar}{iii})$ corresponds to  the value of $T_{\rm e}(\ion{Ar}{iii})$
in units of $10^4$ K. It can be seen from Fig.~\ref{far3o} that there is a good correlation between the temperatures and a fit to the points results in 
\begin{equation}
\label{fitaro}
 t_{\rm e}(\ion{Ar}{iii})=(-0.45\pm0.03) \times t_{3}^{2}  
 + (1.61\pm 0.09) \times t_{3}  - (0.24\pm 0.06).
\end{equation}

The  $\rm Ar^{2+}/H^{+}$  ionic abundance  is calculated by the expression also provided by \citet{Perez-Montero2017}
 
\begin{eqnarray}
\label{eqt4}
 12+\log(\frac{{\rm Ar^{2+}}}{{\rm H^{+}}}) \!\!\!&=&\!\!\! \log \big( \frac{I[\ion{Ar}{iii}]\lambda7135}{I{\rm (H\beta)}}\big)+6.100  \nonumber\\
                                          &&\!\!\!+\frac{0.86}{t_{\rm e}(\ion{Ar}{iii})}-0.404 \times \log t_{\rm e}(\ion{Ar}{iii}). 
\end{eqnarray}
 
\begin{figure}
\centering
\includegraphics[width=0.9\linewidth]{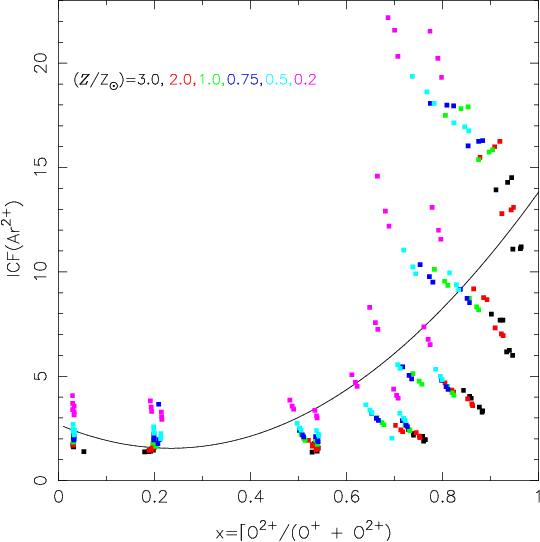}
\caption{ICF for $\rm Ar^{2+}$ (as defined in Eq.~\ref{deficf}) as a function of the $\rm x=[O^{2+}/(O^{+}+O^{2+})]$ abundance ratio.
Points represent results from photoionization models built by \citet{Carvalho2020}.
Model results  assuming different metallicities (in relation to the solar value) are represented by different colours,
as indicated.
The curve represents a fit to all the  points given by the Eq.~\ref{fiticf1}.}
\label{far3o3}
\end{figure}

 To calculate the total argon abundance in relation to
 the  hydrogen (Ar/H) using only the $\rm Ar^{2+}/H^{+}$ abundance
 it is necessary  to consider an ICF in order to take into account the presence of ions with other ionization  levels. In fact, ions such
 as $\rm Ar^{3+}$ and $\rm Ar^{4+}$ have emission lines
 (e.g. [\ion{Ar}{iv}]$\lambda$4740, [\ion{Ar}{v}]$\lambda$7006)
 present in optical AGN spectra (e.g. \citealt{1978ApJ...223...56K, 1992A&A...266..117A}).
 Hitherto, no ICF for the argon abundance determinations
 in AGNs has been proposed in the literature and, in principle,
 due to distinct ionization structure of AGNs, 
 theoretical ICFs proposed by \ion{H}{ii} regions 
 (e.g. \citealt{1985ApJ...291..247M, Izotov2006, Perez-Montero2007, 2020ApJ...893...96B, 2021MNRAS.505.2361A}) can not be applied to this class of  objects.
 In this sense, we used the \citet{Carvalho2020} photoionization models
 to derive an argon ICF,  defined by
 \begin{equation}
 \label{deficf}
     \rm ICF(Ar^{2+})=(Ar/H)/(Ar^{2+}/H^{+})
 \end{equation}
 and in terms of the function $\rm  x=[O^{2+}/(O^{+}+O^{2+})]$
 ionic abundance ratio. In Fig.~\ref{far3o3} the results for ICF($\rm Ar^{2+}$) versus x are shown.
 A fit to the  points  produces the relation  
 \begin{equation}
 \label{fiticf1}
 \rm ICF(Ar^{+})=(a \: x^{2})+ (b\: x) + c,    
 \end{equation}
 where $\rm a=20.88\pm3.80$, $\rm b=-9.74\pm3.64$ and 
$\rm c=2.68\pm0.69$.   A large
scattering of the points can be seen in Fig.~\ref{far3o3}.
We investigated  the source of this scattering   to ascertain whether it is due to variation in the nebular parameters $N_{\rm e}$, $\alpha_{ox}$ 
(slope of the Spectral Energy Distribution, SED\footnote{See \citealt{2021MNRAS.505.2087K} for a detailed description of this SED.}) and   ionization parameter $U$ (not shown). We did not find any dependence  of these parameters with the ICF($\rm Ar^{2+}$)-x relation. However, a clear dependence between ICF($\rm Ar^{2+}$) and  x  on the gas metallicity
is noted in Fig.~\ref{far3o3}, in the sense that for a fixed x value, we derive a higher ICF value with the decrease of the metallicity.
 A  dependence between the metallicity and  ICF-x relations for distinct elements was also derived by
 \citet{Izotov2006} for  SFs.

\begin{figure}
    \centering
    \includegraphics[width=\linewidth]{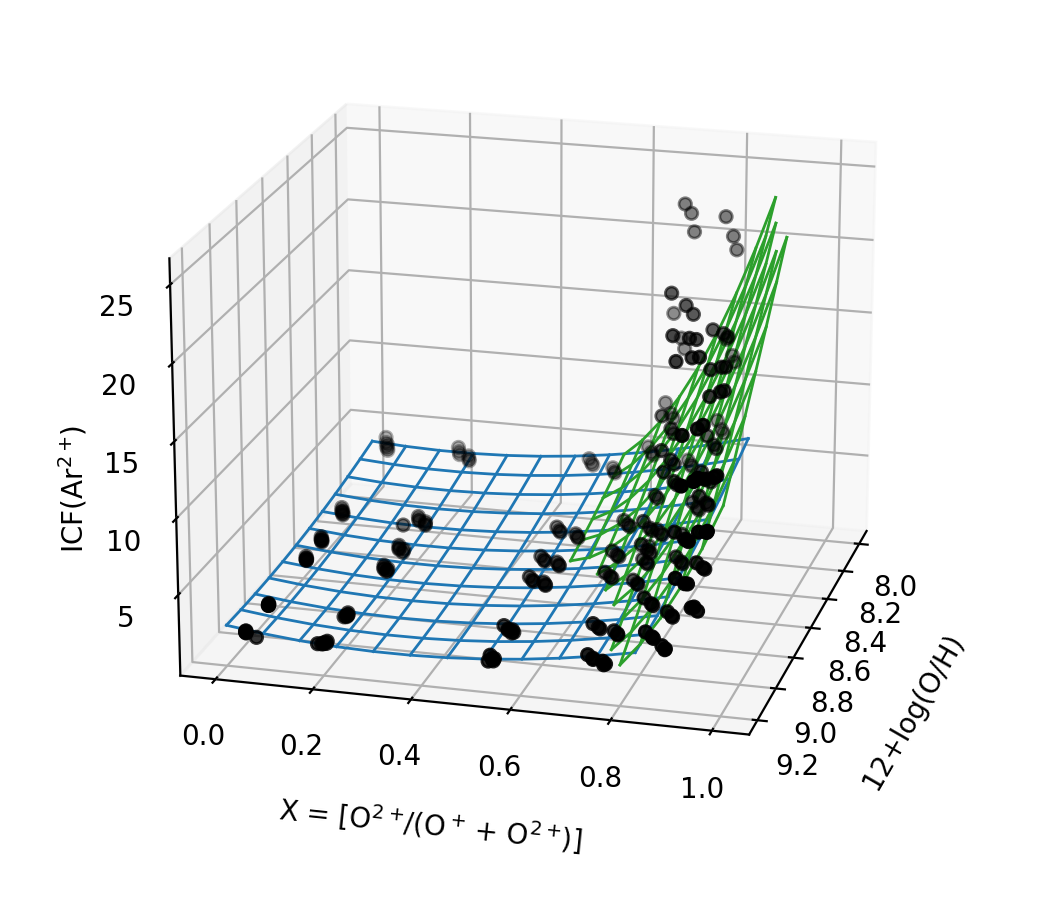}
    \caption{Bi-parametric calibration among the 
$\rm ICF(Ar^{2+})$, $\rm 12+\log(O/H)$ and $\rm x=[O^{2+}/(O^{+}+O^{2+})]$. Points represent the photoionization model results built by \citet{Carvalho2020} while the surfaces  represent the best fit to the points.  The blue surface 
is given by Eq.~\ref{fiticfbi} and valid it is for $\rm x \: < \: 0.6$ while the green surface,  given by Eq.~\ref{fiticfbi2}, is valid for $\rm x \: > \: 0.6$.}
\label{icfbip}
\end{figure}

Thus, in order to obtain a more exact
ICF derivation, we produce a bi-parametric function, i.e.
$\rm ICF(Ar^{2+})=f[12+\log(O/H), x]$. We converted
the metallicity ($Z$) assumed in the photoionization models  built by \citet{Carvalho2020} in oxygen abundance through the expression
\begin{equation}
    12+\log({\rm O/H})= 12+\log[(Z/{\rm Z_{\odot}}) \: \times \:10^{\log(\rm O/H)_{\odot}}],
\end{equation}
where $\log(\rm O/H)_{\odot}=-3.31$ \citep{AllendePrieto2001}.
In Fig.~\ref{icfbip} the bi-parametric calibration  is shown where the points represent the photoionization model results. We fit two functions, one for $\rm x \: < \: 0.6$ represented by a blue surface in Fig.~\ref{icfbip} and given by 
\begin{eqnarray}
\label{fiticfbi}
    {\rm ICF}(\mathrm{Ar^{2+}}) &= (1.51\pm0.3)x^2 - (6.82\pm1.55)y^2 \nonumber\\
                                &- (1.68\pm0.52)xy - (27.18\pm5.19)x \nonumber\\
                                &+ (10.88\pm4.72)y + 124.6\pm22.32
\end{eqnarray}
where $\rm y = 12+\log(\mathrm{O/H})$ and $\rm x=[O^{2+}/(O^{+}+O^{2+})]$
and another one for $\rm x \: > \: 0.6$ represented by the green surface 
and given by 
\begin{eqnarray}
\label{fiticfbi2}
    {\rm ICF}(\mathrm{Ar^{2+}}) &= (1.82\pm142.29)x^2 + (107.86\pm21.07)y^2 \nonumber\\
                                &- (35.61\pm7.13)xy - (14.54\pm33.81)x\nonumber\\
                                &+ (188\pm53.96)y +27.73\pm142.29.
\end{eqnarray}

\noindent The expressions above were applied to derive the total abundance of the argon in relation to the  hydrogen (Ar/H).

The mean errors  for the 12+log(Ar/H) and 12+log(O/H) abundances 
derived for the objects of  our sample are
$\sim$0.25 dex and $\sim$0.13 dex, respectively.
These errors are in order of those derived for O/H abundance determinations in nearby AGNs by 
\citet{2018ApJ...856...46R, 2018ApJ...867...88R, 2021ApJ...910..139R},
who also applied the $T_{\rm e}$-method. 
 However, they 
are somewhat ($\sim$ 0.1 dex) higher than those derived for  disk \ion{H}{ii} regions located in  nearby galaxies (e.g.
\citealt{2020ApJ...893...96B}), because the 
[\ion{O}{iii}]$\lambda$4363 auroral  line is weaker in AGNs than in \ion{H}{ii} regions, resulting in a higher uncertainty in its flux, which in turn  implies a larger uncertainty in the AGN abundance values.

 In Table \ref{tab1}, the SDSS identification of the objects of the sample,  electron
temperature and electron density values, argon ICF values, ionic and total
abundances, redshift, and stellar mass of the hosting galaxies
are listed.

\begin{table*}
\renewcommand{\arraystretch}{0.78}
\setlength{\arrayrulewidth}{0.4pt}
\caption{Results obtained from our sample of objects. Columns: (1) SDSS name, 
(2) $T_{\rm e}$(\ion{O}{iiii}) (in K), (3) $T_{\rm e}$(\ion{A}{iiii}) (in K),
(4) Electron density (in units of $\rm cm^{-3}$), (5) ICF($\rm Ar^{2+}$), (6) $\rm 12+\log(O^{+}/H^{+})$, (7) $\rm 12+\log(O^{2+}/H^{+})$,
(8) $\rm 12+\log(A^{2+}/H^{+})$, (9) 12+log(O/H), (10) 12+log(Ar/H), (11) redshift and (12)  Mass (in units of $10^{7} \: \rm  M_{\odot}$).} 
\label{tab1}
\resizebox{\textwidth}{!}{%
\begin{tabular}{@{}lcccccccccccc@{}}
\hline
(1)  	      &    (2)   &  (3)   &    (4)   &     (5)  &  (6)  &  (7)  &  (8)  &  (9)   &  (10) &   (11)  &  (12)  \\	   
 J101754.72-002811.9  &   14510  & 11481  &    298   &	   2.79 &  8.03 &  7.87 &  5.98 &  8.34  &  6.43 &  0.1817 & 11.13  \\ 
 J101536.21+005459.3  &   18252  & 11994  &    184   &	   2.14 &  8.39 &  7.78 &  5.60 &  8.56  &  5.93 &  0.1202 & 10.71  \\ 
 J104426.16+001707.3  &    8872  &  8343  &    292   &	   2.40 &  8.30 &  8.41 &  6.31 &  8.74  &  6.69 &  0.1502 & 10.71  \\ 
 J105408.69-000111.0  &   16992  & 11971  &    227   &	   2.39 &  8.42 &  7.33 &  5.36 &  8.53  &  5.74 &  0.1081 & 10.21  \\ 
 J111652.97+010615.5  &   16865  &  11952 &	259  &	   2.84 &  8.05 &  7.49 &  5.54 &  8.23  &  6.00 &  0.1309 & 10.70  \\ 
 J114017.31-001543.3  &   13459  &  10677 &	505  &	   2.46 &  8.17 &  7.99 &  6.09 &  8.47  &  6.48 &  0.1245 & 10.62  \\ 
 J113049.84+005346.7  &   19873  &  11827 &	600  &	   2.35 &  8.37 &  7.49 &  5.62 &  8.50  &  5.99 &  0.1043 & 10.70  \\ 
 J113326.76+001443.9  &   14509  &  11386 &	497  &	   2.61 &  8.16 &  7.54 &  5.62 &  8.33  &  6.03 &  0.1143 & 10.48  \\ 
 J115616.76-002221.0  &   15385  &  11724 &	781  &	   2.16 &  8.31 &  7.96 &  6.40 &  8.55  &  6.74 &  0.1092 & 10.74  \\ 
 J122012.58-010531.5  &   11277  &  10039 &    1443  &	   2.18 &  8.39 &  8.43 &  6.24 &  8.79  &  6.58 &  0.1183 & 10.55  \\ 
 J123441.93-010034.7  &   14568  &  11518 &	381  &	   2.52 &  8.15 &  7.75 &  5.55 &  8.37  &  5.95 &  0.0801 & 10.25  \\ 
 J124116.14+004423.0  &   14438  &  11537 &	659  &	   2.24 &  8.26 &  8.02 &  5.90 &  8.54  &  6.26 &  0.0900 & 10.74  \\ 
 J130433.90+000402.9  &   16195  &  11896 &	158  &	   2.56 &  8.19 &  7.53 &  5.92 &  8.35  &  6.33 &  0.2463 & 11.23  \\ 
 J132625.73-002148.6  &   17204  &  11988 &	273  &	   2.89 &  8.10 &  7.25 &  5.53 &  8.24  &  5.99 &  0.1893 & 10.78  \\ 
 J134005.97-010646.4  &   20688  &  11681 &	189  &	   2.35 &  8.48 &  7.29 &  5.43 &  8.59  &  5.81 &  0.1295 & 10.77  \\ 
 J133821.79+002329.2  &   12623  &  10797 &	233  &	   2.60 &  8.17 &  8.25 &  6.14 &  8.59  &  6.55 &  0.1292 & 10.87  \\ 
 J140301.05+005343.5  &   14002  &  11329 &	155  &	   3.03 &  7.94 &  7.65 &  5.51 &  8.20  &  5.99 &  0.1664 & 10.61  \\ 
 J145956.36-002821.5  &   12503  &  10682 &	202  &	   2.57 &  8.13 &  8.01 &  6.14 &  8.45  &  6.55 &  0.1096 & 10.84  \\ 
 J130354.71-030631.8  &   11344  &  10073 &	757  &	   1.84 &  8.53 &  8.40 &  6.28 &  8.85  &  6.55 &  0.0778 & 10.65  \\ 
 J171544.02+600835.4  &   17771  &  12000 &	819  &	   2.14 &  8.38 &  7.79 &  5.59 &  8.56  &  5.92 &  0.1569 & 10.98  \\ 
 J172028.98+584749.6  &   14587  &  11494 &	494  &	   2.92 &  8.03 &  7.42 &  5.29 &  8.21  &  5.75 &  0.1269 & 10.84  \\ 
 J172352.43+582318.5  &   17778  &  11997 &	939  &	   2.91 &  8.07 &  7.31 &  5.20 &  8.22  &  5.66 &  0.0799 & 10.24  \\ 
 J153035.77+001517.7  &   11505  &  10165 &	207  &	   2.28 &  8.24 &  7.93 &  5.87 &  8.50  &  6.23 &  0.0721 & 10.53  \\ 
 J002312.34+003956.3  &   19764  &  11869 &	447  &	   2.06 &  8.54 &  7.73 &  5.67 &  8.68  &  5.98 &  0.0727 & 10.16  \\ 
 J012937.25-003838.6  &   14176  &  11401 &	141  &	   2.57 &  8.16 &  7.60 &  5.51 &  8.34  &  5.92 &  0.1794 & 11.06  \\ 
 J012720.32+010214.6  &    9838  &   9066 &	498  &	   2.27 &  8.41 &  8.55 &  6.36 &  8.87  &  6.72 &  0.1745 & 11.09  \\ 
 J013957.81-004504.2  &   11872  &  10381 &	441  &	   2.74 &  8.11 &  8.18 &  6.40 &  8.52  &  6.83 &  0.1616 & 10.83  \\ 
 J014153.97+010505.4  &   19553  &  11866 &	581  &	   2.23 &  8.40 &  7.62 &  5.33 &  8.55  &  5.68 &  0.1013 & 10.95  \\ 
 J011016.00+150515.9  &   12920  &  10890 &	534  &	   2.28 &  8.27 &  8.19 &  6.10 &  8.61  &  6.46 &  0.0597 & 10.19  \\ 
 J013555.82+143529.6  &   12902  &  10883 &	267  &	   2.52 &  8.14 &  7.86 &  5.68 &  8.40  &  6.08 &  0.0719 & 10.83  \\ 
 J074213.71+391705.3  &   10451  &   9509 &	201  &	   2.40 &  8.22 &  8.19 &  6.19 &  8.59  &  6.57 &  0.0704 &  9.94  \\ 
 J082017.99+465125.3  &   12773  &  10823 &	130  &	   2.13 &  8.46 &  7.70 &  5.64 &  8.61  &  5.97 &  0.0524 & 10.35  \\ 
 J082910.18+504005.7  &   11671  &  10220 &	205  &	   2.00 &  8.44 &  7.92 &  5.91 &  8.64  &  6.21 &  0.0739 &  9.99  \\ 
 J085223.96+531550.6  &   17610  &  11995 &	524  &	   2.52 &  8.27 &  7.40 &  5.54 &  8.41  &  5.94 &  0.1280 & 10.78  \\ 
 J095123.44+581621.2  &   18905  &  11952 &	140  &	   2.45 &  8.48 &  6.96 &  5.12 &  8.58  &  5.51 &  0.1486 & 10.91  \\ 
 J033923.14-054841.5  &   12841  &  10856 &	331  &	   2.54 &  8.16 &  8.13 &  5.86 &  8.52  &  6.27 &  0.0848 & 10.26  \\ 
 J091605.16+002030.3  &   12682  &  10791 &	482  &	   1.83 &  8.53 &  8.14 &  5.77 &  8.76  &  6.03 &  0.1434 & 10.97  \\ 
 J093509.12+002557.4  &   15921  &  11833 &	551  &	   2.70 &  8.11 &  7.52 &  5.24 &  8.29  &  5.68 &  0.1512 & 10.78  \\ 
 J100013.84+624703.4  &   10996  &   9868 &	488  &	   2.55 &  8.15 &  8.09 &  5.96 &  8.50  &  6.37 &  0.1145 & 10.33  \\ 
 J102039.81+642435.8  &   15695  &  11779 &	172  &	   2.29 &  8.25 &  7.86 &  5.47 &  8.47  &  5.83 &  0.1223 & 10.74  \\ 
 J095759.45+022810.5  &   12988  &  10905 &	325  &	   2.35 &  8.23 &  8.11 &  6.12 &  8.55  &  6.49 &  0.1194 & 10.56  \\ 
 J100921.26+013334.5  &   13197  &  11029 &	735  &	   2.20 &  8.32 &  8.27 &  6.04 &  8.68  &  6.39 &  0.1437 & 10.72  \\ 
 J112748.89+020302.6  &   16053  &  11845 &	451  &	   2.25 &  8.30 &  7.76 &  5.53 &  8.49  &  5.88 &  0.1267 & 10.80  \\ 
 J114304.62+013946.2  &   10858  &   9810 &	197  &	   2.03 &  8.38 &  8.04 &  5.94 &  8.62  &  6.25 &  0.0928 & 10.84  \\ 
 J114029.55+022744.6  &   20626  &  11680 &	375  &	   2.48 &  8.28 &  7.44 &  5.23 &  8.42  &  5.62 &  0.1230 & 10.63  \\ 
 J115854.96+033254.9  &   15449  &  11741 &	373  &	   2.27 &  8.36 &  7.63 &  5.41 &  8.51  &  5.77 &  0.0841 & 10.51  \\ 
 J125503.63+012233.7  &   13282  &  11026 &	731  &	   2.09 &  8.37 &  8.25 &  5.94 &  8.69  &  6.26 &  0.1642 & 10.93  \\ 
 J125209.68+021558.0  &   15928  &  11815 &	200  &	   2.53 &  8.28 &  7.38 &  5.26 &  8.41  &  5.66 &  0.2064 & 10.95  \\ 
 J130220.35+024048.8  &   13312  &  11045 &	710  &	   2.33 &  8.23 &  7.85 &  5.98 &  8.46  &  6.35 &  0.1766 & 10.93  \\ 
 J134959.37+030058.0  &   14538  &  11545 &	346  &	   2.25 &  8.26 &  7.97 &  6.17 &  8.52  &  6.52 &  0.1097 & 10.35  \\ 
 J140231.58+021546.3  &   11109  &   9934 &	263  &	   2.03 &  8.44 &  8.42 &  6.07 &  8.81  &  6.38 &  0.1797 & 11.06  \\ 
 J143214.54+023228.5  &   15605  &  11787 &	439  &	   2.31 &  8.26 &  7.77 &  5.58 &  8.46  &  5.95 &  0.1123 & 10.91  \\ 
 J074257.23+333217.9  &   14084  &  11333 &	366  &	   2.77 &  8.05 &  7.66 &  5.53 &  8.27  &  5.97 &  0.1474 & 10.71  \\ 
 J090246.69+520932.8  &   13797  &  11254 &	290  &	   2.34 &  8.22 &  8.02 &  5.70 &  8.51  &  6.07 &  0.1375 & 10.95  \\ 
 J141530.97+035916.6  &   16651  &  11928 &	463  &	   2.41 &  8.40 &  7.31 &  5.56 &  8.52  &  5.95 &  0.0805 & 10.49  \\ 
 J141351.75+042208.9  &   16889  &  11957 &	293  &	   2.47 &  8.23 &  7.57 &  5.41 &  8.40  &  5.80 &  0.1449 & 10.60  \\ 
 J144925.29+044157.2  &    9467  &   8810 &	126  &	   1.91 &  8.46 &  8.15 &  6.14 &  8.71  &  6.42 &  0.0824 & 10.04  \\ 
 J151244.15+042848.3  &   14481  &  11439 &	575  &	   2.37 &  8.23 &  7.73 &  5.66 &  8.43  &  6.04 &  0.0796 & 10.34  \\ 
 J155404.39+545708.2  &   12320  &  10609 &	147  &	   2.48 &  8.20 &  8.20 &  6.00 &  8.58  &  6.39 &  0.0457 &  9.91  \\ 
 J164938.71+420658.4  &   14941  &  11624 &	345  &	   2.34 &  8.30 &  7.63 &  5.61 &  8.46  &  5.98 &  0.1503 & 10.58  \\ 
 J165944.29+392846.1  &   16801  &  11947 &	285  &	   2.32 &  8.32 &  7.62 &  5.78 &  8.48  &  6.15 &  0.0818 & 10.09  \\ 
 J100602.50+071131.8  &   14649  &  11535 &    1588  &	   1.87 &  8.64 &  7.95 &  5.76 &  8.80  &  6.04 &  0.1205 & 11.19  \\ 
 J163344.99+372335.1  &   20228  &  11735 &	771  &	   2.50 &  8.21 &  7.57 &  5.54 &  8.38  &  5.94 &  0.1748 & 10.80  \\ 
 J125558.75+291459.4  &   11324  &  10055 &	366  &	   2.00 &  8.44 &  8.36 &  6.21 &  8.78  &  6.51 &  0.0681 &  9.91  \\     	      
\hline
\end{tabular}%
}

\end{table*}

\section{Results and Discussion} 
\label{resc}

\begin{figure}
\centering
\includegraphics[width=0.9\linewidth]{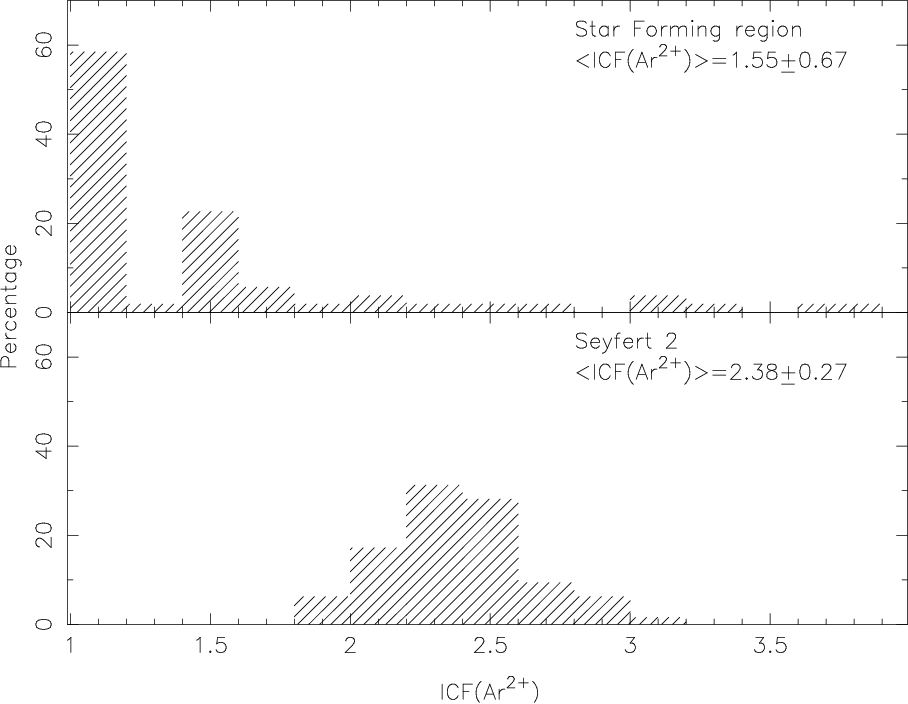}
\caption{Bottom panel: distribution of the ICF values for the argon from our sample of Seyfert~2 (see Sect.~\ref{obs}) derived by using  Eqs.~\ref{fiticfbi} and \ref{fiticfbi2}.
The averaged value is indicated. Top panel: Same as the bottom panel but for
ICFs of SFs (\ion{H}{ii} regions and \ion{H}{ii} galaxies) derived 
by \citet{2020ApJ...893...96B, 2015ApJ...806...16B}, \citet{2016ApJ...830....4C, 2015ApJ...808...42C}, \citet{2009ApJ...700..309B}, \citet{Hagele2008}, and \citet{2004ApJ...614..698L}.}
\label{hicf}
\end{figure}

 The total argon abundance relative to hydrogen (Ar/H) in gaseous nebulae  has been derived mainly through measurements of the  [\ion{Ar}{iii}]$\lambda7135$/H$\beta$ 
and the assumptions considered to convert this line ratio into the $\rm Ar^{2+}/H^{+}$ ionic abundance as well as by using
an expression for the  ICF($\rm Ar^{2+}$) (e.g. \citealt{Kennicutt2003}).
 In some few cases, it has been also possible to measure the [\ion{Ar}{iv}]$\lambda$4740 and [\ion{Ar}{v}]$\lambda$7006 emission lines and
deriving the $\rm Ar^{3+}/H^{+}$ and  $\rm Ar^{4+}/H^{+}$ ionic abundances, respectively (e.g. \citealt{1978ApJ...223...56K, 1981ApJ...246..434F,  2018A&A...615A..29P}), producing a more exact Ar/H determination, i.e.
\begin{equation}
\rm \frac{Ar}{H}=ICF(Ar)\times \left[ \frac{Ar^{2+}}{H^{+}}+\frac{Ar^{3+}}{H^{+}}+\frac{Ar^{4+}}{H^{+}}\right]
\end{equation}
(e.g. \citealt{2005MNRAS.362..424W}). However, the [\ion{Ar}{iv}]$\lambda$4740 and [\ion{Ar}{v}]$\lambda$7006 emission
lines are weak in most part of low ionization objects, therefore, it is most convenient  to use only the strong [\ion{Ar}{iii}]$\lambda7135$ (e.g. \citealt{Kennicutt2003, 2012MNRAS.427.1463Z}). In our case, the emission
line measurements made available by the MPA/JHU group only  allowed us to derive $\rm Ar^{2+}/H^{+}$ ionic abundance and then apply the ICF (Eq.~\ref{fiticfbi}) to obtain the total argon abundance (Ar/H).
The application of the $T_{\rm e}$-method  to AGNs was recently proposed by \citet{Dors2020b} and the ICF for the argon, proposed in this study, hitherto, is the first one for this class of objects. Thus, it is worthwhile to analyse the derived ICF  values for our sample as well as to compare it with other proposed derivations for gaseous nebulae. In Fig.~\ref{hicf}, bottom panel, the distribution of the argon ICF values
derived from our sample through Eq.~\ref{fiticfbi} is shown. We derived
ICF values ranging from $\sim1.8$ to $\sim 3$, with an averaged value 
of $2.38\pm0.27$. Also in Fig.~\ref{hicf}, top panel, the argon ICF distribution
for 53  SFs (49 disk \ion{H}{ii} regions and 4 \ion{H}{ii} galaxies) compiled
from the literature is shown. We notice that majority of SFs 
present lower ICFs in comparison with those derived from our sample
of Seyfert~2 nuclei, implying that $\sim 90\%$ of the SFs have
$\rm ICF(Ar^{2+})\: \la \: 2$ and an average value of about 1.5.
These results point to the known fact that the gas phase in
Seyfert~2 has a higher excitation degree than  SFs.

\begin{figure}
\centering
\includegraphics[width=0.9\linewidth]{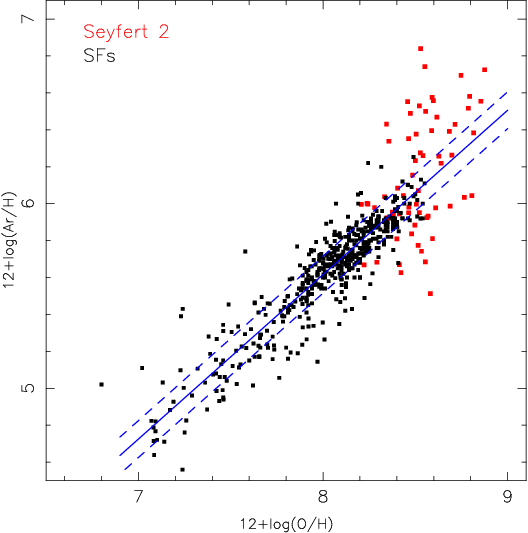}
\caption{Total argon abundance [in units of 12+log(Ar/H)]  versus the
oxygen abundance [in units of 12+log(O/H)]. Red points represent abundance
results from our sample of Seyfert~2. Black points represent
abundance results from \ion{H}{ii} galaxies obtained through $T_{\rm e}$-method
by \citet{Izotov2006} and \citet{Hagele2008}.
The solid line represents a fit to the points (Eq.~\ref{eqaroh})
while the dashed lines represent the uncertainty of $\pm 0.1$ dex in \ion{H}{ii} region abundance estimations (e.g. \citealt{Kennicutt2003}).}
\label{farvox}
\end{figure}

Regarding  abundance estimates, our results for Seyfert~2 nuclei  can be
used to verify a possible secondary stellar production of the argon
at the very high metallicity regime $\rm [12+log(O/H) \: \ga \: 8.8]$,
 because the maximum  value derived in most part of SFs through the $T_{\rm e}$-method
is in the order of 12+log(O/H)$\sim8.7$ (e.g. \citealt{Kennicutt2003, 2020ApJ...893...96B, 2020A&A...634A.107Y}). In Fig.~\ref{farvox}, the total argon abundance
[in units of 12+log(Ar/H)]  versus the
oxygen abundance [in units of 12+log(O/H)] from our sample of Seyfert~2 nuclei is shown.
Also in this figure, abundance results based on $T_{\rm e}$-method for   galaxy nuclei with star formation  derived by
\citet{Izotov2006} (whose observational data were also taken from SDSS; \citealt{2005AJ....129.1755A}) as well as  results for \ion{H}{ii} galaxies
obtained by  \citet{Hagele2008} are shown. In this case, we  chose to compare our results only with those  from star-forming galaxies
(excluding disk \ion{H}{ii} regions) because
 SFs are subject to  similar
physical processes like those in
AGNs. The following are few examples which underscore such similarities.
\begin{enumerate}
    \item  SFs and AGNs  can present  gas outflows (e.g. \citealt{2012MNRAS.427..968H, 2013MNRAS.436.2929H, 2017MNRAS.469.4831C, 2014ApJ...780L..24R, 2021MNRAS.501L..54R, 2021MNRAS.506.2950R, 2021arXiv210802334C}).
    \item  In  the case of spiral  galaxies the presence of bars might produce a falling of gas into the central regions  in both objects (e.g. \citealt{1992MNRAS.259..345A}).
\end{enumerate}
  These processes can regulate the star formation  as well as modify the  gas enrichment (e.g. \citealt{2010MNRAS.406.2325O, 2017MNRAS.466.1213K, 2017MNRAS.471..144S, 2020MNRAS.495.2564J}) 
   and could not be present in  isolated
   \ion{H}{ii} regions.
From Fig.~\ref{farvox}, we observe a clear linear relation between Ar/H and O/H indicating 
a primary production of the argon in a wide metallicity regime.
A fit to the points (considering Seyfert~2 and SFs) produces the expression
\begin{equation}
\label{eqaroh}
\rm 12+log(Ar/H)= [(0.89\pm0.01) \times 12+log(O/H)]-(1.50\pm0.15),
\end{equation}
represented in Fig.~\ref{farvox} by a solid blue line. The dashed lines in this figure represent the uncertainty of $\pm 0.1$ dex in abundance estimations (e.g. \citealt{Kennicutt2003}). 
It is noteworthy to point out from Fig.~\ref{farvox} that, for very high O/H values,   a slight deviation of the points representing  AGNs results from Eq.~\ref{eqaroh}. This can be an indication
of a secondary argon production at very high metallicity
regime. In view of this, we tested several
polynomial fittings considering different  limits for oxygen abundance  looking for a transition from primary to secondary Ar production, 
but no satisfactory solution was obtained. Therefore, we emphasize that
our  results obtained from AGNs, when combined with those from star-forming galaxies, indicate that the argon has a primary stellar production in a wide metallicity range.

 The  heavy elements in AGNs are mainly produced  by stellar evolution in the ISM
located at few  kpc away from the nuclei (e.g. \citealt{1993A&A...277..397B, 2002AJ....123.1381E, 2007MNRAS.382..251D, 2008A&A...482...59D, 2008AJ....135..479B, 2009MNRAS.393..783R, 2013MNRAS.432..810H, 2015MNRAS.451.3173A, 2016MNRAS.461.4192R, 2020A&A...639A..96P, 2021ApJ...908..155M}) and  then transported
to the supermassive black hole. Additional metal enrichment by stars can also be
obtained by two ways:
\begin{enumerate}
    \item $in$ $situ$ star formation   embedded in the  thin AGN accretion disk
(e.g. \citealt{1999Ap&SS.265..501C, 2004ApJ...608..108G, 2008A&A...477..419C, 2011ApJ...739....3W, 2012ApJ...749..168M,  2020MNRAS.498.3452D, 2021ApJ...910...94C}) and by
\item capture of stars orbiting the central regions of galaxies
(e.g. \citealt{1991MNRAS.250..505S, 1993ApJ...409..592A}).
\end{enumerate}
\noindent Both processes can produce different chemical evolution of AGNs in comparison with that of SFs.
For instance, the maximum oxygen  abundance in the centres of  most luminous star-forming galaxies has been found to be 
12+log(O/H)$\sim8.9$ (e.g. \citealt{2007MNRAS.376..353P}) while Seyfert~2 nuclei can reach up to $\sim9.2$ dex
(e.g. \citealt{Dors2020b, 2021arXiv210711606D}), i.e.
there is an additional enrichment in Seyferts in comparison with metal rich \ion{H}{ii} regions. Moreover, recently, \citet{2021arXiv210904596A} found that
Ne/H abundance in a small sample of Seyfert 2s is nearly 2 times higher than those in SFs. 
Thus, determining metal abundance in AGNs, in this case argon abundance, can produce constraints in studies on star nucleosynthesis in the very high metallicity regime and in different boundary conditions than those in SFs.  

\begin{figure}
    \centering
    \includegraphics[width=\linewidth]{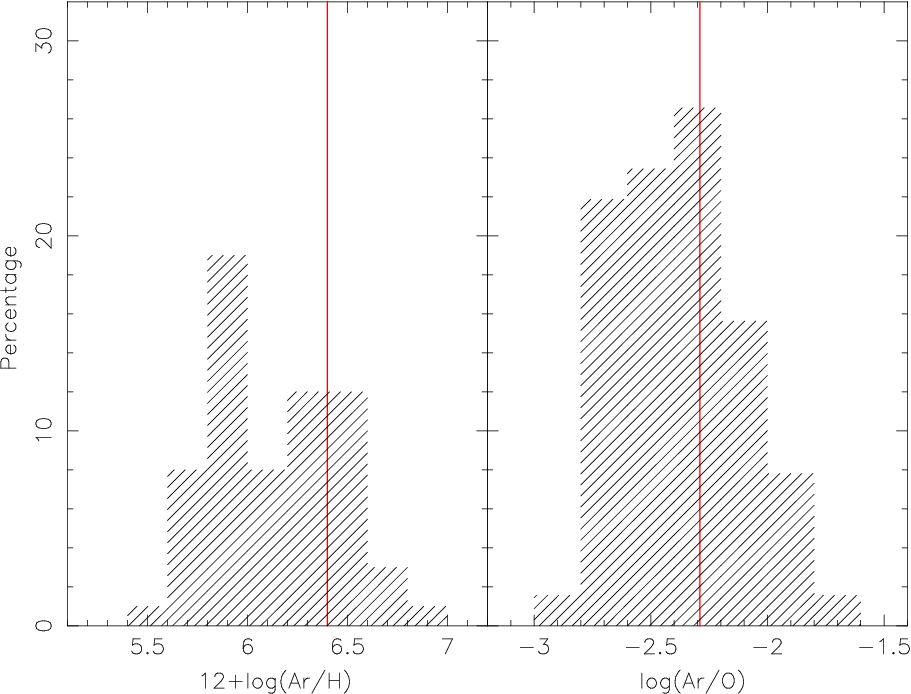}
    \caption{Left panel: distribution of argon abundance [in units of 12+log(Ar/H)] for our sample
    of Seyfert~2 nuclei. The red line represents the solar argon abundance [$\rm 12+log(Ar/H)_{\odot}=6.40$] derived by \citet{1998SSRv...85..161G}. Right panel: same as the left panel
    but for log(Ar/O).  The red line represent the solar argon abundance [$\rm \log(Ar/O)_{\odot}=-2.29$] derived by \citet{1998SSRv...85..161G} and \citet{AllendePrieto2001}.}
    \label{histoaro}
\end{figure}

As a result, it is important to verify the range of Ar/H abundance in AGNs, since these objects
present, in general, higher metallicity (or O/H abundance) than star-forming regions (e.g. \citealt{groves2006emission, Dors2020b}). 
In Fig.~\ref{histoaro}, the distributions of argon abundance [in units of 12+log(Ar/H)]  and log(Ar/O) for our sample and the solar value for these abundances are shown. In Table~\ref{tabraro} the minimum, maximum and the mean  abundance values, taken from Fig.~\ref{histoaro},  as well as the oxygen abundance values (not shown in Fig.~\ref{histoaro}) are listed. We also listed in Table~\ref{tabraro} the minimum,  maximum and mean abundance values ($W_{\rm v}$) in relation  to the corresponding solar value, defined by 
\begin{equation}
\label{maxx}
W_{\rm v}^{\rm X}=\rm X(v)/X_{\odot},    
\end{equation}
where X is a given abundance ratio. The solar argon and oxygen abundances are those derived by \citet{1998SSRv...85..161G} and \citet{AllendePrieto2001}, respectively. 
It can be inferred from  Table~\ref{tabraro} that the maximum  Ar/H and Ar/O abundance ratios in our sample of AGNs are higher than the solar values, i.e. $\sim 2.7$
and $\sim 4.0$ times the solar values, respectively.
However, the  $W_{\rm max.}^{\rm Ar/H}$ and $W_{\rm max.}^{\rm Ar/O}$ can even be higher than the ones found in this work due to the sample considered.
Actually, the  $W_{\rm max.}^{\rm O/H}$ derived for our sample is
about 1.5 times the solar value.
 \citet{Dors2020b} found  $W_{\rm max.}^{\rm O/H}\sim 3.2$ after considering a larger sample of SDSS Seyfert~2 nuclei  without adopting the [\ion{Ar}{iii}] emission line presence as a selection criterion. Thus, this disparity  indicates the probable
existence  of oversolar Ar/H and Ar/O values  higher than those listed in Table~\ref{tabraro}.

\begin{table*}
\caption{Minimum, maximum and the mean  abundance ratio values 
for our sample (see Sect.~\ref{obs})
derived through  $T_{\rm e}$-method adapted for AGNs
(see Sect.~\ref{secteme}). The abundance values 
are obtained from the  distributions presented in Fig.~\ref{histoaro}. The $W_{\rm min.}^{\rm X}$, $W_{\rm max.}^{\rm X}$ and $W_{\rm mean}^{\rm X}$ are calculated from Eq.~\ref{maxx}
and represent abundance values relative to the solar values.
The solar argon and oxygen abundances
 considered were those derived by \citet{1998SSRv...85..161G} and \citet{AllendePrieto2001}.}
\label{tabraro}
\begin{tabular}{lrrrrrrr}
\hline
Abundance ratio     &        Min.          &       Max.          &  Mean              & $W_{\rm min.}^{\rm X}$&  $W_{\rm max.}^{\rm X}$ &  $W_{\rm mean}^{\rm X}$ 	  \\
\hline              
12+log(Ar/H)        &           5.51       &     6.84            & $ 6.14 \pm 0.32$   &     0.13              &  2.75                   &  $0.54^{+0.60}_{-0.27}$	  \\[5pt]
log(Ar/O)           &         $-3.06$      &     $-1.68$         & $-2.37  \pm 0.28$  &     0.16              &  3.98                   &  $0.97^{+0.61}_{-0.53}$	  \\[5pt]
12+log(O/H)         &           8.20       &     8.87            & $ 8.52 \pm 0.16$   &     0.30              &  1.51                   &  $0.67^{+0.30}_{-0.20}$	   \\
\hline
\end{tabular}
\end{table*}

\begin{figure}
    \centering
    \includegraphics[width=\linewidth]{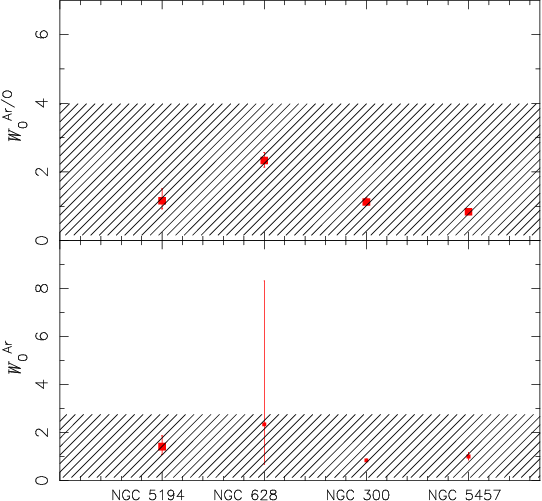}
    \caption{Comparison between argon abundance in relation
to the solar value for $W_{0}^{\rm Ar}=\rm (Ar)/(Ar)_{\odot}$ (bottom panel) and $W_{0}^{\rm Ar/O}=\rm (Ar/O)/(Ar/O)_{\odot}$ (top panel). Points represent
estimates from radial abundance  extrapolations to the central parts
 (galactocentric distance $R=0$) from the four spiral galaxies listed in Table~\ref{tab2a}   and indicated in the
x-axis}. The hatched areas represent 
the range of abundance ratio values derived for our sample of Seyfert~2 
listed in Table~\ref{tabraro}.
\label{compwaof}
\end{figure}

In order to compare the abundance range of Ar/H and Ar/O derived through
the new methodology  from the $T_{\rm e}$-method for AGNs with those  obtained from other methods, we compiled
from the literature Ar/H and Ar/O radial gradients derived in spiral 
galaxies and  calculated by the standard $T_{\rm e}$-method. We extrapolated these radial gradients to the central part of the galaxies (i.e. galactocentric distance $R=0$). This methodology makes it possible to infer indirect and independent values of abundances in the nuclei of spiral galaxies (e.g. \citealt{1992MNRAS.259..121V, 1998AJ....116.2805V, 2004A&A...425..849P, 2019MNRAS.483.1901Z}).
As usual, we assume that the Ar/H and Ar/O radial abundance gradients are represented by 
\begin{equation}
\label{eqY2}
Y^{\rm X}  = Y{_0}^{\rm X} + [grad\,Y^{\rm X} \times R (\rm kpc)],
\end{equation} 
where $Y^{\rm X}$ represents the value of the abundance ratio X [12+log(Ar/H) or log(Ar/O)],
 $Y_{0}^{\rm X}$ is the extrapolated value of X  to the galactic center, i.e. at radial distance $R=0$,
and  $grad\,Y^{\rm X}$ is the slope of the distribution expressed in   units of $\rm dex \: kpc^{-1}$.
As pointed out by \citet{2004A&A...425..849P}, the  accuracy of  radial  abundance  gradient determinations is defined not only by the large number of objects considered
but also by the distribution  of these objects along the galactic radius.  Under this supposition, we take into consideration published data from the literature for X abundance values of  \ion{H}{ii} regions derived by using the $T_{\rm e}$-method and located at galactocentric distances in spiral galaxies within the range 
$0.2 \:  \:  \la \: (R/R_{25}) \: \la \: 1$, where $R$ is the galactocentric distance and $R_{25}$ is the $B$-band isophote at a surface brightness of 25 mag arcsec$^{-2}$.  
It was possible to obtain the  Ar/H and Ar/O radial  gradients
in four spiral galaxies. In Table~\ref{tab2a}, the identification of each galaxy,
the number ($N$) of \ion{H}{ii} regions considered in deriving  the  radial gradients,
the $Y{_0}^{\rm X}$ and $grad\,Y^{\rm X}$ values as well as references to the original  studies from which the data were obtained are listed. Also in Table~\ref{tab2a}, the $Y_{0}^{\rm X}$ 
in relation to the solar value,  defined by
\begin{equation}
\label{neonsolar}
 W_{0}^{X}= Y_{0}^{\rm X}/(\rm X)_{\odot}    
\end{equation}
is listed. In Fig.~\ref{compwaof}, the $W_{0}^{\rm X}$ values
for the four spiral galaxies, listed in Table~\ref{tab2a},
are represented by red points and compared with those for the range
of values (hatched areas) derived for our sample and listed in Table~\ref{tabraro}. It can be seen that our range of abundance estimates  are in consonance with the values obtained by independent abundance estimates from the radial gradients. Obviously, a more exact comparison  would be obtained if both   Ar/H and O/H abundances in the AGNs and the  radial abundances  of these elements in the host galaxies
were determined, such as  studies carried out for the oxygen by
\citet{1998AJ....115..909S} and \citet{Dors2015}.

\begin{table*}
\caption{Parameters of the  radial abundance gradients derived for the Ar/H and Ar/O abundance ratios in a sample of spiral galaxies.
$N$ represents the number of \ion{H}{ii} regions considered in the estimations of the
gradients. $Y{_0}$,  $grad\,Y$ and $W_{0}$ are defined in Eqs.~\ref{eqY2}
and \ref{neonsolar}. In the last column, the original works from which the 
radial gradients  were compiled are listed.}
\label{tab2a}
\begin{tabular}{lrcrrrrrrrr}
\hline
Object    & $N$  & &   $Y{_0}^{\rm Ar/H}$        & $grad\,Y^{\rm Ar/H}$      & $W_{0}^{\rm Ar}$  & & $Y{_0}^{\rm Ar/O}$         & $grad\,Y^{\rm Ar/O}$	         &$ W_{0}^{\rm Ar/O}$ &    Reference                    \\
\hline 
NGC\,5194 & 28   & &  $6.55\pm 0.12$&  $-0.015\pm 0.026$ &	$1.41^{+0.45}_{-0.33}$             &  & $-2.22 \pm 0.11$  & $+0.013 \pm 0.023$   &    $1.16^{+0.35}_{-0.24}$               & \citet{2015ApJ...808...42C}      \\[5pt] 
NGC\,628  & 45   & &  $6.77 \pm 0.55$&  $-0.070 \pm 0.009$ &	$2.34^{+5.97}_{-1.66}$             &  & $-1.92 \pm 0.04$  & $-0.037 \pm 0.006$   &     $2.33^{+0.24}_{-0.19}$              & \citet{2015ApJ...806...16B}      \\[5pt] 
NGC\,300  & 28   & &  $6.32\pm 0.03$&  $-0.106 \pm 0.011$ &	$0.84^{+0.05}_{-0.03}$             &  & $-2.24 \pm 0.03$  & $-0.028 \pm 0.009$   &     $1.12^{+0.12}_{-0.07}$              & \citet{2009ApJ...700..309B}      \\[5pt]
NGC\,5457 & 16   & &  $6.40 \pm 0.07$&  $-0.020 \pm 0.003$ &	$1.00^{+0.17}_{-0.14}$             &  & $-2.36 \pm 0.01$  & $-0.008 \pm 0.002$   &    $0.84^{+0.03}_{-0.01}$               & \citet{Kennicutt2003}            \\
\hline
\end{tabular}
\end{table*}

\begin{figure}
    \centering
    \includegraphics[width=\linewidth]{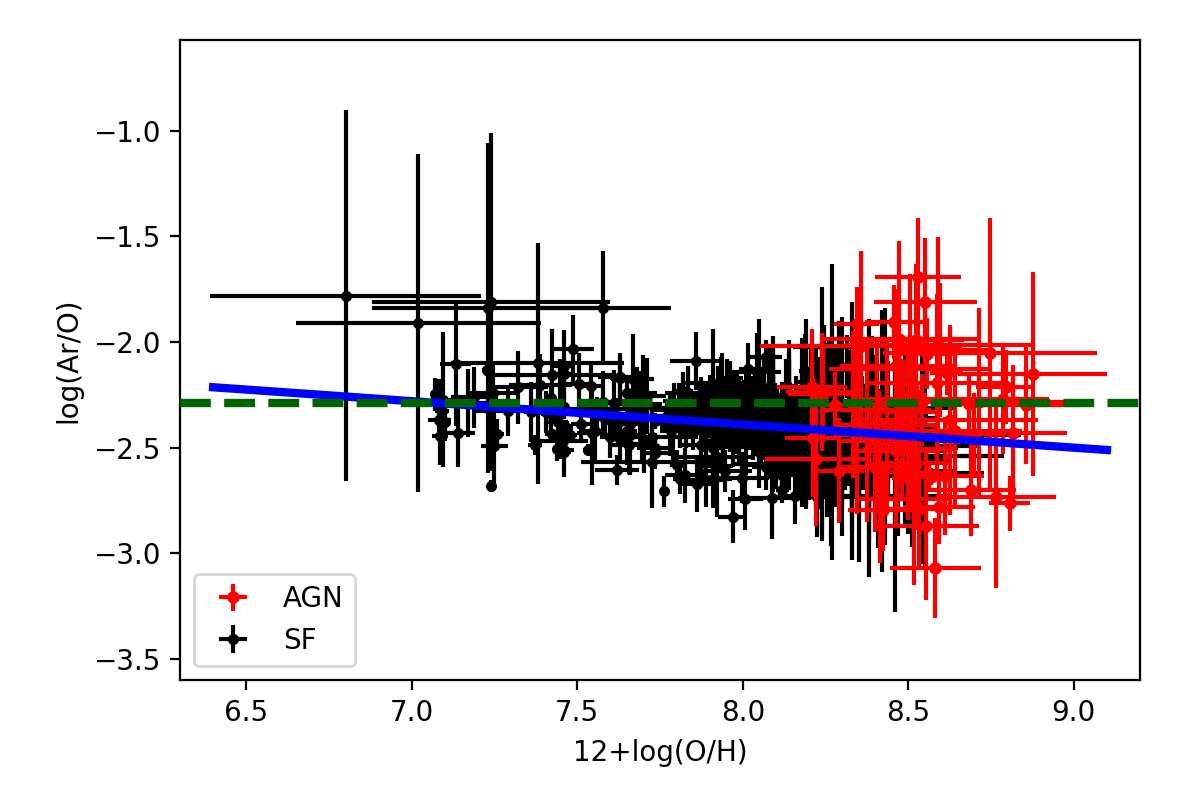}
    \caption{Relation between log(Ar/O) and 12+log(O/H). 
Red points represent  estimations for our sample of Seyfert~2 
while black points represent compiled estimations for \ion{H}{ii} galaxies
derived by \citet{Izotov2006} and \citet{Hagele2008}.
The blue solid line represents the linear regression to the points 
given by Eq.~\ref{ffit}. The dashed green line represents
the solar value, log(Ar/O)$_{\odot}=-2.29$, derived by \citet{1998SSRv...85..161G} and \citet{AllendePrieto2001}.
}
    \label{figaovo}
\end{figure}

Finally, in Fig.~\ref{figaovo}, our Seyfert~2 estimates (red points) for (Ar/O) versus (O/H) abundances and those calculated through the $T_{\rm e}$-method for
 star-forming nuclei
 by \citet{Izotov2006} and  \ion{H}{ii} galaxies by \citet{Hagele2008} are shown.
As for the neon and sulphur (see \citealt{2013MNRAS.432.2512D, 2016MNRAS.456.4407D} and references therein) several studies on  
(Ar/O) versus (O/H) have yielded conflicting results.
For instance, \citet{Izotov2006}, by correcting the depletion of oxygen onto dust grains (which is in order of $\sim 0.1 \: \rm  dex$) found a negative slope
for Ar/O versus O/H  for the range $\rm 7.1 \: \la \: [12+log(O/H)] \: \la \: 8.6$, which correspond to the metallicity range 
$\rm 0.03 \: \la \: (Z/\rm Z_{\odot}) \: \la 0.7$. This negative slope is strongly influenced 
by  estimations in objects with very low metallicity   $\rm [\rm 12+\log(O/H)] \: \la \: 7.5]$,  in which oversolar Ar/O abundance values  are derived. Although this result of a negative slope is in consonance with the one found by \citet{Perez-Montero2007}, who  considered
\ion{H}{ii} galaxies  and \ion{H}{ii} regions estimates, these authors derived
solar Ar/O abundance ratios for objects at low metallicity regime and
subsolar values for the ones at high metallicity regime.
Finally, \citet{2020ApJ...893...96B}, who took into account  the
$T_{\rm e}$-method abundance estimates for 190 disk \ion{H}{ii} regions, derived  sub-solar  Ar/O  abundance ratio for some few  \ion{H}{ii} regions with low metallicity  $[\rm 12+\log(O/H) \: \la \: 8.0$]. In principle, a possible explanation for the discrepancies above is the use of different argon ICFs by the authors rather than  the consideration of distinct samples. In any case,  it can be seen from  Fig.~\ref{figaovo} that, a slight decrease of Ar/O with the increase of O/H, when our estimates are combined with those for 
\ion{H}{ii} galaxies. It is worth to mention that, for   $[\rm 12+\log(O/H) \: \ga \: 8.0$] our Ar/O estimates are in consonance with those   derived for \ion{H}{ii} galaxies. A  linear regression to the points produces   
\begin{equation}
\label{ffit}
\rm log(Ar/O) = (-0.11 \pm 0.02)\:x - (1.51 \pm 0.15),    
\end{equation}
where x=log(O/H).

\section{Conclusions}
\label{conc}
For the fist time, quantitative argon abundances, based on $T_{\rm e}$-method adapted for AGNs, is derived  for the narrow regions of Seyfert~2 nuclei. In view of this,
we compiled  from Sloan Digital Sky Survey Data Release 7 (SDSS-DR7) optical narrow emission line fluxes for 64 Seyfert 2 galaxies in the local universe ($z \: < \: 0.25$) and calculated the abundance of the argon and oxygen in relation to
hydrogen (Ar/H, O/H).
The total argon abundance relative to hydrogen (Ar/H)  was derived for each 
object of the sample through the  $ \rm  Ar^{2+}/H^+$ 
ionic abundance and by using a theoretical expression for the Ionization Correction Factor (ICF) 
obtained from photoionization model results built with the {\sc Cloudy} code. These results from the models
were also used to derive an appropriate temperature for the $\rm Ar^{2+}$  [$T_{\rm e}(\ion{Ar}{iii})$]
which can be derived by its dependence on the temperature for $\rm O^{2+}$ 
[$T_{\rm e}(\ion{O}{iii})$]
calculated by using the observational 
[\ion{O}{iii}]($\lambda$4959+$\lambda$5007)/$\lambda$4363 line ratio.  
We obtained the following conclusions:

\begin{enumerate}
\item The equality between the temperatures $T_{\rm e}(\ion{S}{iii})$=$T_{\rm e}(\ion{Ar}{iii})$,
usually assumed in  abundance studies of star-forming regions, is not valid for Seyfert~2  since
the nebular gas region occupied by $\rm Ar^{2+}$ tends to have a lower temperature than the nebular gas region occupied by  $\rm S^{2+}$. \\

\item A bi-parametric expression for the ICF($\rm Ar^{2+}$) as  function of the 
$\rm x=[O^{2+}/(O^{+}+O^{2+})]$ abundance ratio and  the oxygen abundance [in units of 12+log(O/H)]
is proposed
to derive the total argon abundance.\\

\item For the range of oxygen abundance $\rm 8.0 \: \la \: [12+\log(O/H)] \: \la \: 9.0$ or metallicity
$0.20 \: \la \: (Z/\rm \rm Z_{\odot}) \: \la \:2.0$, we found that our sample 
of Seyfert~2 present A/H abundances ranging from  $\sim 0.1$ to $\sim 3$ times the argon solar value, indicating that most of the objects ($\sim 75\%)$ have subsolar argon abundance.\\

\item The  range of Ar/H and Ar/O abundance values obtained for our sample are in consonance with those estimated from extrapolations to the central parts of  radial abundance gradients derived in the disk of four spiral galaxies.\\

\item We found a slight tendency of Ar/O abundance ratio  decreases with O/H.\\

\item Finally, the Ar/O abundance values for our 
  sample of Seyfert~2 are in consonance with estimations for  \ion{H}{ii} galaxies
  $[\rm 12+\log(O/H) \: \ga \: 8.0$], indicating that there is not an over  enrichment of argon in AGNs, at least for the metallicity range considered.

\end{enumerate}

\section*{Acknowledgements}
AFM  gratefully acknowledges support from Coordenação de Aperfeiçoamento de Pessoal de Nível Superior (CAPES). OLD is grateful 
to  Funda\c c\~ao de Amparo \`a Pesquisa do Estado de S\~ao Paulo (FAPESP) and Conselho Nacional
de Desenvolvimento Cient\'{\i}fico e Tecnol\'ogico (CNPq).

\section{DATA AVAILABILITY}
The data underlying this article will be shared on reasonable request
with the corresponding author.

%%%%%%%%%%%%%%%%%%%%%%%%%%%%%%%%%%%%%%%%%%%%%%%%%%



\bibliographystyle{mnras}
\bibliography{artigo} % if your bibtex file is called example.bib

%\begin{thebibliography}{99}
%\end{thebibliography}

%%%%%%%%%%%%%%%%%%%%%%%%%%%%%%%%%%%%%%%%%%%%%%%%%%

%%%%%%%%%%%%%%%%% APPENDICES %%%%%%%%%%%%%%%%%%%%%

%\begin{figure*}
%    \centering
%    \includegraphics[width=\linewidth]{view1.jpg}
%    \caption{Caption}
%    \label{fig:my_label}
%\end{figure*}

%%%%%%%%%%%%%%%%%%%%%%%%%%%%%%%%%%%%%%%%%%%%%%%%%%

% Don't change these lines
\bsp	% typesetting comment
\label{lastpage}
\end{document}